\documentclass[]{aastex63}
\pdfoutput=1

\usepackage{mathtools}
\shorttitle{Solar Flare Prediction with Machine Learning}
\shortauthors{Wang et al.}

\begin{document}

\title{Predicting solar flares with machine learning: investigating solar cycle
dependence}

\correspondingauthor{Xiantong Wang}
\email{xtwang@umich.edu}

\author[0000-0002-8963-7432]{Xiantong Wang}
\affiliation{Department of Climate and Space Sciences and Engineering, University of Michigan \\
Ann Arbor, MI, USA}

\author[0000-0002-9516-8134]{Yang Chen}
\affiliation{Department of Statistics, University of Michigan\\
Ann Arbor, MI, USA}

\author[0000-0002-5654-9823]{Gabor Toth}
\affiliation{Department of Climate and Space Sciences and Engineering, University of Michigan \\
Ann Arbor, MI, USA}

\author[0000-0003-0472-9408]{Ward B. Manchester}
\affiliation{Department of Climate and Space Sciences and Engineering, University of Michigan \\
Ann Arbor, MI, USA}

\author[0000-0001-9360-4951]{Tamas I. Gombosi}
\affiliation{Department of Climate and Space Sciences and Engineering, University of Michigan \\
Ann Arbor, MI, USA}

\author{Alfred O. Hero}
\affiliation{Department of Electrical Engineering and Computer Science, University of Michigan\\
Ann Arbor, MI, USA}

\author{Zhenbang Jiao}
\affiliation{Department of Statistics, University of Michigan\\
Ann Arbor, MI, USA}

\author{Hu Sun}
\affiliation{Department of Statistics, University of Michigan\\
Ann Arbor, MI, USA}

\author[0000-0002-9672-3873]{Meng Jin}
\affiliation{Lockheed Martin Solar and Astrophysics Laboratory, Palo Alto, California, USA}
\affiliation{SETI Institute, Mountain View, CA 94043, USA}

\author[0000-0002-0671-689X]{Yang Liu}
\affiliation{Hansen Experimental Physics Laboratory, Stanford University, Stanford, CA 94305, USA}



\begin{abstract}

A deep learning network, Long-Short Term Memory (LSTM), is used to predict whether an active region (AR) will produce a flare of class $\Gamma$ in the next 24 hours. We consider $\Gamma$ being $\ge M$ (strong flare), $\ge C$ (medium flare) and $\ge A$ (any flare) class. The essence of using LSTM, which is a recurrent neural network, is its capability to capture temporal information of the data samples. The input features are time sequences of 20 magnetic parameters from the Space-weather HMI Active Region Patches (SHARPs). We analyze active regions from June 2010 to Dec 2018 and their associated flares identified in the (Geostationary Operational Environmental Satellite) GOES X-ray flare catalogs. Our results (\romannumeral1) produce skill scores consistent with recently published results using LSTMs and are better than the previous results using single time input. (\romannumeral2) The skill scores from the model show statistically significant variation when different years of data are chosen for training and testing. In particular, the years 2015 to 2018 have better True Skill Statistic (TSS) and Heidke Skill Scores (HSS) for predicting $\ge C$ medium flares than the years 2011 to 2014 when the difference in flare occurrence rates is properly taken into account. 

\end{abstract}



\section{Introduction} \label{sec:intro}
Solar flares are energetic eruptions of electromagnetic radiation from the Sun lasting from minutes to hours. 
The terrestrial impact of small flares is limited, but strong flares have a significant on the upper atmosphere. Increased ionization affects the total electron content, which in turn affects radio wave propagation and global positioning system (GPS) accuracy.  Ionospheric heating causes the atmosphere to expand, increasing the mass density and increasing drag on satellites altering their orbits. Strong flare are also often accompanied with coronal mass ejections (CMEs) that can cause substantial impact on the Earth environment. Therefore, it is very worthwhile to improve the prediction of solar flares, especially larger ones. 
During solar cycle 24, nearly 800 M or X flares were observed. While posing a significant threat, the rareness of extreme events and the complexity of the flares makes solar flare time and intensity predictions a very challenging task. 

Although the triggering mechanism of solar flares and the factors determining the solar flare strength are far from being well understood, it is shown by multiple studies that solar flares are caused by the sudden release of free energy brought by magnetic reconnection in the coronal field.  What has come to be know as the standard model for flares and CMEs \citep{carmichael1964, sturrock1966, hirayama1974theoretical, kopp1976} (also called the CSHKP model), involves the rise of sheared core or flux rope that results in magnetic reconnection in the surrounding arcade structure.  Several variations of this model have been developed, which incorporate different initiation mechanisms \citep{masuda1994loop, forbes1996reconnection, manchester2003, torok2004}. A number of review papers summarize these works and many others \citep{green2018}. 

Since the photospheric magnetic field drives the coronal field, it is possible that the evolution patterns of the photospheric magnetic field may serve as indicators of the triggering process of flares and CMEs. Those features include the size of the active regions (AR), the integrated magnetic flux, the integrated current helicity, the magnetic field gradient measurements, the shear angle of the magnetic field structure and so on. The Helioseismic and Magnetic Imager (HMI) on the Solar Dynamics Observatory (SDO) satellite, launched a decade ago, has been providing high-cadence high-resolution photospheric vector magnetic field observations starting from 2010. The Space-weather HMI Active Region Patches, a.k.a. SHARPs \citep{2014SoPh..289.3549B}, contain time series data localized to individual active regions (ARs) with many pre-calculated quantities based on the AR magnetic. We will use these SHARP quantities to train our machine learning model. 

Machine learning, a sub-field of artificial intelligence, utilizes past data as a "learning context" for the computer program that allow it to make predictions of the future state of the system. The advent of the Solar Heliophysics Observatory (SOHO)/Michelson Doppler Imager (MDI) and SDO/HMI missions provided sufficient data for machine learning algorithms to predict solar activities. 
At first, the line-of-sight (LOS) component of the photospheric magnetic field measured by the MDI instrument was used by several groups to forecast solar flares using machine learning algorithms (\cite{2013SoPh..283..157A}, \cite{2018ApJ...856....7H}, \cite{2009SoPh..254..101S}, \cite{2009SoPh..255...91Y}, \cite{2010RAA....10..785Y}). The support vector machine (SVM) algorithm was used by \cite{2015ApJ...812...51B} for a classification task on time series of MDI data from 2000 to 2010. 
However, the LOS magnetic field component does not include all the magnetic field information, so later studies used the vector magnetic field data once it became available from the HMI instrument. \cite{2015ApJ...798..135B} used the SVM trained with SHARP parameters for active region classification tasks. 
\cite{2018ApJ...858..113N} built a residual deep neural network using not only the parameterized photospheric magnetograms but also using chromospheric images. \cite{2018SoPh..293...48J} used observations from photosphere, chromosphere, transition region, and corona as input of the machine learning algorithm, which gave a comparable result to the works done by \cite{2015ApJ...798..135B} and \cite{2018ApJ...858..113N}. 

The machine learning algorithms in these studies mentioned so far do not fully utilize the time dependence of the input. Among various kinds of machine learning algorithms, recurrent neural networks (RNNs) are suitable to analyze time series input. 
Long-short term memory networks (LSTMs) \citep{hochreiter1997long, gers1999learning}, a particular kind of RNNs, have succeeded in many sequence classification and prediction tasks, including speech recognition, time-series forecasting, handwriting recognition and so on \citep{hastie2005elements, graves2013speech}. 
Most recently, \cite{2019ApJ...877..121L}, \cite{chen2019lstm} and \cite{jiao2019solar} used LSTMs on SHARP parameters, which achieved better performance for predicting solar flares compared to previous works. \cite{sun2019interpreting} identifies key signals for strong flares from SHARP parameters using time-series clustering on LSTMs predictions.

In this paper, we apply the LSTM algorithm on the SHARP parameters from SDO/HMI vector magnetic field to predict the maximum solar flare class produced by an active region in the next 24 hours. The inputs are 48-hour time series of SHARP parameters with 12-minute cadence. The observations of ARs are time sequences, hence LSTMs are suitable for this kind of input. First, our results show consistency with recently published work by \cite{2019ApJ...877..121L} that also uses the LSTM algorithm. Second, we also find that the skill scores vary substantially when using different years of ARs in the training and testing set. This indicates that data samples should be carefully chosen for the model evaluation. It is also of interest to understand in what respect these years differ in the solar cycle from each other that may make the solar flare prediction less or more successful. 

The rest of this paper is organized as follows. Section \ref{sec:details} describes how we collect data and build the training and testing sets. Section \ref{sec:arch} describes the LSTM architecture we are using in this work. Section \ref{sec:model_evaluation} explains the metrics used to evaluate the model performance. Section \ref{sec:results} shows the results of this study and compares them with previous work. The solar cycle dependence of the prediction skills are also presented in this section. Section \ref{sec:discussion_conclusion} describes our conclusions.

\section{Details of the Data Preparation}\label{sec:details}

\subsection{Dataset}\label{subsec:construct_dataset}
We use SHARP summary parameters as the input data of the prediction model. The Space-weather HMI Active Region Patches (SHARPs) is a data product derived from vector magnetograms taken from the \textit{Helioseismic and Magnetic Imager} (HMI) onboard the \textit{Solar Dynamics Observatory} (SDO) \citep{2014SoPh..289.3549B}. 
The summary parameters are calculated based on the \textit{HMI Active Region Patches} (HARPs), which are rectangular boxes surrounding the active regions that are moving with the solar rotation and the evolution of the active regions. Table \ref{tab:sharp_description} lists the 20 key parameters used in this work. 
The SHARP summary parameters are downloaded for all active regions from the \href{http://jsoc.stanford.edu}{Joint Science Operations Center} (\texttt{jsoc.stanford.edu}) from 2011 to 2018. 
The solar flare events are identified from the NOAA Geostationary Operational Environmental Satellites (GOES) flare list \citep{1994SoPh..154..275G}. 
In the GOES flare list, flare events are listed with class, start, end, and peak intensity times of each event. 
The peak time of the flare events are assigned as the "event time" when constructing the data samples. 
The number of active regions and flare events in different years are summarized in Table \ref{tab:solar_flare_number}. 
Note that C flares outnumber the A and B flares suggesting that most of the A and B flares are missed when their relatively weak signal falls below the X-ray background.

The input of our model are the time sequences of the SHARP summary parameters. To guarantee the quality of the data, some time sequences are dropped, especially when the active regions are at the limb. The criteria for dropping unqualified time sequences are as follows:
\begin{enumerate}
    \item In order to avoid projection effects, the longitude of the HARP region center is within the range of $\pm 68^\circ$ from the central meridian, 
    \item The fraction of missing frames in a time sequence has to be less than $5\%$,
    \item Two time sequences are separated by one hour.
\end{enumerate}
The target value (or label) of each sequence is the maximum flare class produced by the active region in the next 24h after the end time of the sequence. 
The NOAA active region number is used to match the HARP and AR numbers in the GOES flare list. 
However, while GOES flares are identified strictly with NOAA ARs, we note that a single AR may be split among multiple HARPs or that a HARP may contain multiple ARs. 
Consequently, we find that $20\%$ of HARPs have this mismatch issue, which may lead to a potential error when we assign the maximum flare classes to the time sequences for flares may be missed or improperly attributed to the HARP.

Because the various SHARP features have different scales and units, the original data samples are normalized before input into the machine learning model: let $z_i^n$ denote the normalized value of the $i^{th}$ feature in the $n^{th}$ data sample, then

\begin{equation}
    z_i^n = \frac{v_i^n - \mu_i}{\sigma_i},
\end{equation}
where $v_i^n$ is the original value of feature $i$ in data sample $n$, while $\mu_i$ and $\sigma_i$ are the mean and standard deviation of the feature $i$ calculated from the entire dataset, respectively.

\begin{deluxetable}{ccccccc}
    \tablenum{1}
    \tablecaption{Number of active regions and flares of different classes observed each year. \label{tab:solar_flare_number}}
    \tablewidth{0pt}
    \tablehead{
    \colhead{Year} & \colhead{ARs} & \colhead{A} & \colhead{B} & \colhead{C} & \colhead{M} & \colhead{X}
    }
    \startdata
    2011 & 168 & 1  & 665 & 1002 & 106 & 9 \\      
    2012 & 168 & 0  & 475 & 1115 & 124 & 7 \\
    2013 & 183 & 0  & 469 & 1197 & 97  & 12\\
    2014 & 194 & 0  & 184 & 1627 & 194 & 16\\
    2015 & 143 & 0  & 446 & 1274 & 128 & 2 \\
    2016 & 109 & 0  & 757 & 294  & 15  & 0 \\
    2017 & 52  & 0  & 620 & 229  & 37  & 4 \\
    2018 & 21  & 5  & 255 & 12   & 0   & 0 \\
    \hline
    \enddata
\end{deluxetable}

\begin{deluxetable*}{ll}
    \tablenum{2}
    \tablecaption{List of SHARP parameters and brief descriptions\label{tab:sharp_description}}
    \tablewidth{0pt}
    \tablehead{
    \colhead{Parameter} & \colhead{Description} 
    }
    \startdata
    USFLUX & Total unsigned flux in Maxwells \\
    MEANGAM & Mean inclination angle, gamma, in degrees \\
    MEANGBT & Mean value of the total field gradient, in Gauss/Mm  \\
    MEANGBZ & Mean value of the vertical field gradient, in Gauss/Mm  \\
    MEANGBH & Mean value of the horizontal field gradient, in Gauss/Mm  \\
    MEANJZD & Mean vertical current density, in mA/$m^2$  \\
    TOTUSJZ & Total unsigned vertical current, in Amperes  \\
    MEANALP & Total twist parameter, alpha, in 1/Mm  \\
    MEANJZH & Mean current helicity in $G^2$/m  \\
    TOTUSJH & Total unsigned current helicity in $G^2$/m  \\
    ABSNJZH & Absolute value of the net current helicity in $G^2$/m \\
    SAVNCPP & Sum of the Absolute Value of the Net Currents Per Polarity in Amperes \\
    MEANPOT & Mean photospheric excess magnetic energy density in ergs per cubic centimeter \\
    TOTPOT & Total photospheric magnetic energy density in ergs per cubic centimeter \\
    MEANSHR & Mean shear angle (measured using $B_{total}$) in degrees \\
    SHRGT45 & Percentage of pixels with a mean shear angle greater than 45 degrees in percent \\
    SIZE & Projected area of patch on image in micro-hemisphere\\
    SIZE\_ACR & Projected area of active pixels on image in micro-hemisphere\\
    NACR & Number of active pixels in patch\\
    NPIX & Number of pixels within the patch\\
    \enddata
\end{deluxetable*}

\subsection{Training/Testing splitting}\label{subsec:train_test_split}
In order to assess the performance of the machine learning algorithms properly, we need to split the samples (time sequences of SHARP summary parameters and corresponding maximum flare classes) into a training set and a testing set. The training set is used for training the machine learning model while the testing set is for assessing the prediction capability of the model. In the training process, the model learns from the input data and adjusts its parameters to fit the ground truth. Both variable selection and parameter estimation are included in this process. The samples in the testing set should be totally separated from the training set, otherwise there will be an artificial gain of performance since the information in the training set is leaked to the testing set \citep{kaufman2012leakage, schutt2013doing}. Hence, separating the samples based on active regions is necessary to guarantee that sequences from one active region will not occur in both training and testing sets simultaneously. All the training/testing splitting in this paper are conducted based on HARPs.

\section{Architecture of Machine Learning Model}\label{sec:arch}
The Recurrent Neural Network (RNN) is a category of neural networks which can make use of sequential information \citep{pearlmutter1989learning}. 
This architecture is naturally used in solar flare prediction since the active regions evolve with time and the occurrence of the solar flares is most likely related to the time-dependent evolution of active regions. 
RNNs are called recurrent because they perform the same task for every input from the sequence, but the output depends on the previous computations. 
Among various RNN structures, the Long Short Term Memory (LSTM) network is one of the most commonly used type of RNNs \citep{hochreiter1997long}. 
LSTM networks are explicitly designed to avoid the long-term dependency problem, which is a major shortcoming for simpler RNNs. 
The key to LSTMs is a new cell state variable in the network, which is passed through the whole chain with only minor linear interactions. 
This allows the information at a much earlier time to effect the results, which mimics a `long-term' memory.  

\begin{figure*}[htb!]
    \centering
    \includegraphics[width=0.85\textwidth]{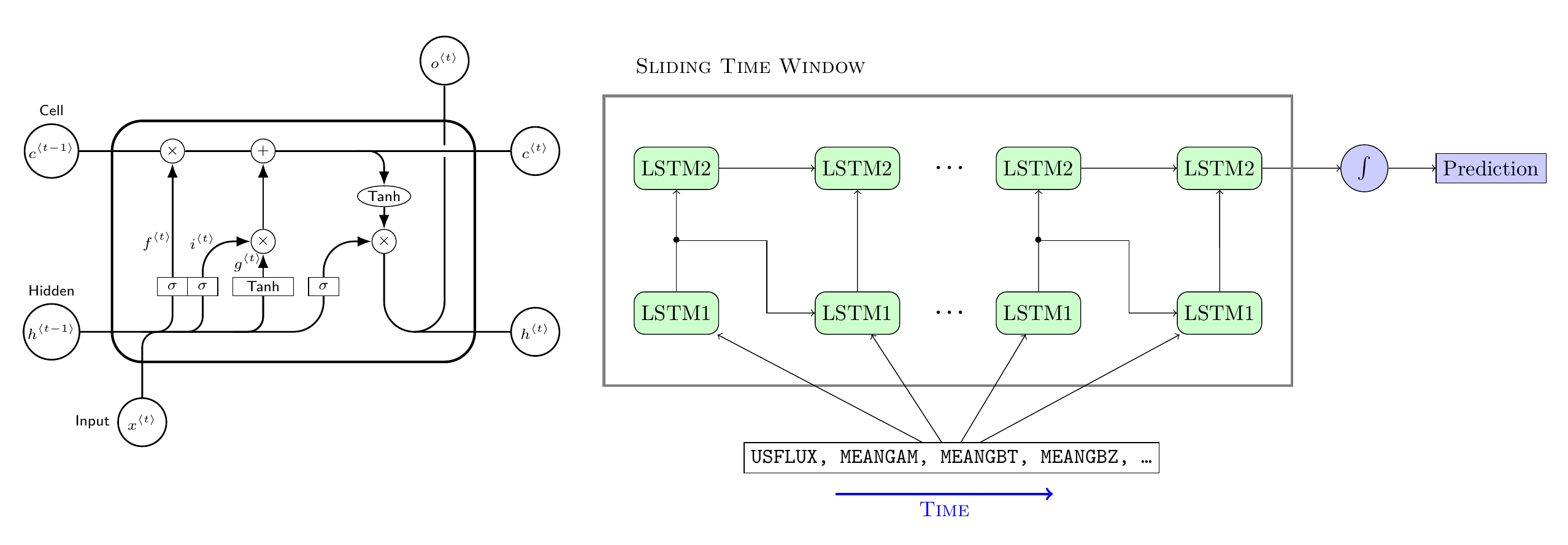}
    \caption{The detailed structure of the LSTM cell (left) and the LSTM network (right).}\label{fig:lstm_arch}
\end{figure*}

The structure of the LSTM unit and the LSTM network used in this work is shown in Figure \ref{fig:lstm_arch}. 
The left panel shows a single LSTM unit. 
Each unit takes an input vector $x^{<t>}$ consisting of the input features at a certain time point $t$, the hidden state $h^{<t-1>}$ and the cell state $c^{<t-1>}$ are from the previous LSTM unit. 
The right panel shows the structure of the two-layer LSTM network. 
The LSTM units in the second layer are the same as in the first layer, but their input vectors $x^{<t>}_2$ are the output vectors $o^{<t>}_1$ from the first layer LSTM units. 
The relationships between the unit input, output and internal states are given by the following equations:

\begin{subequations}\label{eqn:lstm_unit}
\begin{align}
    i^{<t>} &= \sigma(W_{ii} x^{<t>} + b_{ii} + W_{hi} h^{<t-1>} + b_{hi}) \\
    f^{<t>} &= \sigma(W_{if} x^{<t>} + b_{if} + W_{hf} h^{<t-1>} + b_{hf}) \\
    g^{<t>} &= \tanh(W_{ig} x^{<t>} + b_{ig} + W_{hg} h^{<t-1>} + b_{hg}) \\
    o^{<t>} &= \sigma(W_{io} x^{<t>} + b_{io} + W_{ho} h^{<t-1>} + b_{ho}) \\
    c^{<t>} &= f^{<t>} * c^{<t-1>} + i^{<t>} * g^{<t>} \\
    h^{<t>} &= o^{<t>} * \tanh(c^{<t>})
\end{align}
\end{subequations}

Here $x^{<t>} \in \mathbb{R}^d$ is the input vector to the LSTM unit and $d$ is the number of features. 
$i^{<t>}$, $f^{<t>}$ and $o^{<t>}\in \mathbb{R}^h$ are the activation vectors of the input gates, forget gates and output gates, respectively. 
$g^{<t>} \in \mathbb{R}^h$ is an activation vector from the $\tanh$ function. 
$h$ is the hidden dimension of the LSTM unit, which is a hyperparameter in the model that reflects the model complexity and we use 16 in this work. $c^{<t>} \in \mathbb{R}^h$ is the cell state vector and there is only linear relationship between the output and input cell states in a single LSTM unit. The cell state and hidden state vectors are passed to the next LSTM unit in the same layer. The output vectors in the first layer are taken as input in the second layer. The output vector of the last LSTM unit in the second layer is multiplied by a $h \time 1$ vector and passed to a sigmoid function to generate the final prediction value.
The $\tanh$ and the sigmoid function 

\begin{equation}\label{eqn:sigmoid}
    \sigma(x) = \frac{1}{1 + e^{-x}}
\end{equation}
introduce the non-linearity into the neural network.
The $W_{i\_} \in \mathbb{R}^{h \times d}$ weight matrices are applied to the input vectors and $W_{h\_} \in \mathbb{R}^{h \times h}$ are applied to the gate activation vectors. $b \in \mathbb{R}^h$ in the equation are the bias vectors. The weight matrices and bias vectors are trainable parameters, which are determined during the training process.

The "training" in the machine learning is essentially an optimization process for an objective function, also known as the loss function. The loss function measures the difference between the model prediction and the ground truth. An optimization algorithm is used to minimize the loss function so that the trainable parameters in the model can "encode" some knowledge from the data samples.  The prediction can finally be reduced to a binary classification task: according to the given input, will this AR produce a flare of class $\Gamma$ in the next 24 hours. The model will generate a prediction score in the last layer in the network and if this score is larger than a threshold, then the model will make a positive prediction. In our work, the last layer is a sigmoid function and the output from a sigmoid function is either close to 0 or 1, so the threshold for binary classification is set to be 0.5. The "Binary Cross Entropy" is typically used as the loss function for binary classification problems. However, this loss function can fail if one category of samples is dominating the entire data set. Constantly predicting the dominant category in the testing set can result in a small value of the loss function but the model has no predictive skill in this case. Large energetic solar flares are extremely rare events so that the dataset we are using is highly unbalanced. To solve this issue, we used "Binary Cross Entropy with Logits Loss" in this work which is defined as:

\begin{equation}\label{eqn:bcewithlogits}
    L = \frac{1}{N}\sum_{n=1}^N -[p_c y_n \log{\sigma(\hat{y}_n)} + (1-y_n)\log(1 - \sigma(\hat{y}_n))]
\end{equation}
Here $N$ is the number of samples in the training or testing set, $y_n$ is the target value and $\hat{y}_n$ is the model output. The coefficient $p_c$ and the use of the sigmoid function distinguish this loss function relative to the simple binary cross entropy loss function. The sigmoid function improves the numerical stability in the optimization process, while $p_c$ is set to the ratio of negative and positive samples to balance the contributions of the two terms in the sum. In this work, this value is calculated from the training set. 

\section{Model Evaluation}\label{sec:model_evaluation}

The four quantities TN, TP, FN, and FP refer to the number of True Negative, True Positive, False Negative and False Positive predictions, respectively. These four numbers can be combined to calculate the Precision, the Recall (also known as Probability of Detection, POD), the False Alarm Rate (FAR), the True Skill Statistic (TSS), the Heidke Skill Score (HSS), and the Accuracy (ACC) defined as

\begin{subequations}
    \begin{align}
    \rm Precision &= \rm \frac{TP}{TP + FP} \\
    \rm POD &= \rm Recall =\rm \frac{TP}{TP + FN}\\
    \rm FAR &= \rm \frac{FP}{FP + TN} \\
    \rm TSS &= \rm \frac{TP}{TP + FN} - \frac{FP}{FP + TN} = POD - FAR\\
    \rm HSS &= \rm \frac{2(TP \cdot TN - FP \cdot FN)}{(TP + FN)(FN + TN) + (TP + FP)(FP + TN)} \\ 
    \rm ACC &= \rm \frac{TP + TN}{TP + FP + FN + TN} \\
    \end{align}
\end{subequations}
We will use these quantities to evaluate the model performance. 
The Precision and POD evaluate the model's ability to identify positive events (1 being perfect, 0 being worst), while FAR tests the model for correctly identifying negative events (0 being perfect, 1 being worst). 
The Accuracy, TSS and HSS evaluate the overall skill with the maximum value 1 being the perfect score. 
As we will see, the various skill scores have very different dependence on the fraction of positive and negative events in the training and testing sets. 
For an imbalanced dataset, the Accuracy becomes less meaningful because the model's output will be dominated by the majority of the dataset.
Artificial inflation will be caused to the POD (or the FAR) if the model is assigning all testing samples to be positive (or negative). 
However, in both cases, the model will not have any useful prediction skills. TSS approaches the POD when the forecasting is dominated by correct forecasts of non-occurrence which is the case for solar flare events. 
A high TSS value therefore may not really mean that the prediction is reliable, as there can be many false alarms relative to the number of true predictions. 
HSS is superior to the TSS in this situation, because it produces 0 value for a model that predicts a random number with the correct occurrence rate, and positive HSS means that the model is better than that. 
However, HSS is sensitive to the ratio of positive versus negative events, which means that the same model can produce very different HSS values depending on the selected data set (for example solar maximum versus solar minimum). 
In other words, HSS can be scaled up if there are more positive samples in the testing set \citep{doswell1990summary}. 
Hence there is no single skill score that can properly evaluate the forecasting performance, and one needs to be careful when models are compared. 
There are only four independent values (TN, TP, FN, and FP) and only their three ratios truly matter. 
This means that any three independent values defined above can be used.
In practice, we concentrate on the POD, FAR and HSS values, as these provide complete and intuitive information about the model's performance. The TSS, while useful, is not an independent skill score, as it is simply the difference of POD and FAR. 

\section{Results}\label{sec:results}

\subsection{Training Process}\label{subsec:train_proc}
The LSTM network is implemented in Python with the PyTorch package. 
PyTorch is originally a tensor calculation package for GPU and the auto-gradient feature \citep{paszke2017automatic} makes it suitable for machine learning tasks. 
A minibatch strategy \citep{li2014efficient, bottou2018optimization} is used for faster convergence during back-propagation. The Adam optimizer \citep{kingma2014adam}is used with the learning rate set to 0.001 and the other parameters are $\beta_1=0.9$ and $\beta_2=0.999$. 
The batch size is 1000. 
The model is trained for multiple epochs on the training set. 
In each epoch, the model goes through the training samples once. 
The model is trained for 6 epochs to generate the results presented in Table \ref{tab:comp_skill_scores}, Figure \ref{fig:comp_year} and Figure \ref{fig:comp_base_model}. 
We found no statistically significant improvement in the performance after six epochs. 
To account for the randomness due to the order of training samples and the initial values of the trainable parameters, we perform 20 independent runs with different random seeds to get evaluate robustness.

\subsection{Skill scores for solar flare prediction}

\begin{figure}[htb!]
    \centering
    \includegraphics[width=0.45\textwidth]{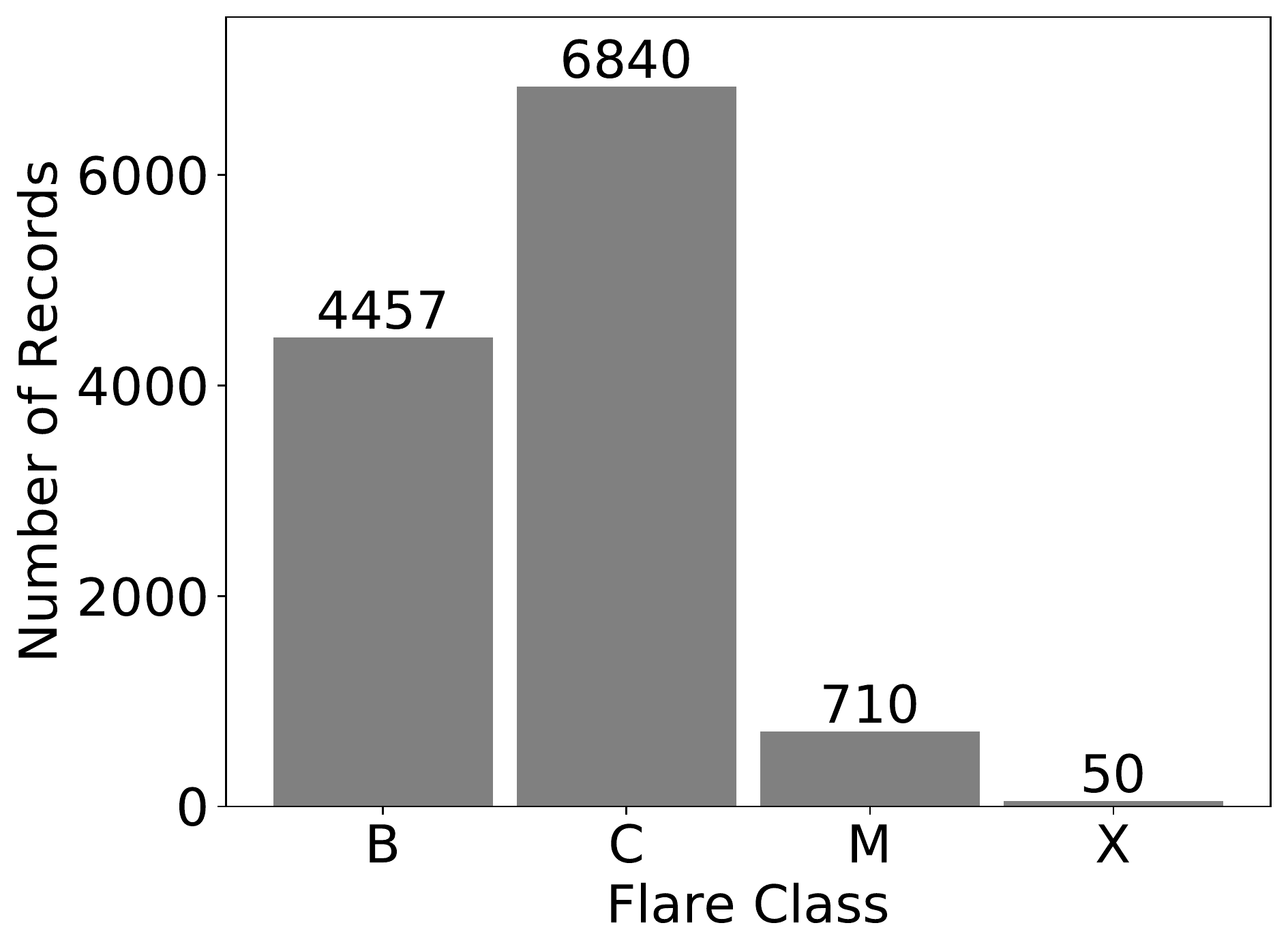}
    \caption{The distribution of number of recorded flares for classes $B, C, M, X$ from 2011 to 2018 in the GOES data set. Note that the number of $B$ flares is much smaller than the number of $C$ flares.} \label{fig:missing_B}
\end{figure}

The mean values of the skill scores for the 20 runs are reported in Table \ref{tab:comp_skill_scores} together with results obtained by earlier work for comparison.
From Table \ref{tab:comp_skill_scores}, the skill scores for predicting $\ge C$ flares are larger than those predicting $\ge M$ flares in all models. 
This is expected, because there are many more C flares than M and X flares, which helps the machine learning algorithm to get better.
However, the skill scores for predicting any flares ($\ge A$) are also smaller than those predicting $\ge C$ flares. 
The reason for this reduced performance is that A and B flares are not properly observed.
The number of flares in different energy classes roughly obey a power-law distribution (see \cite{lu1991avalanches} and Figure \ref{fig:flare_count_hist}), thus a large number of class $B$ flares are missing in the GOES flare records (see Figure \ref{fig:missing_B}). 
The X-ray emission of many $B$ flares can fall below the background emission level once an active region heats up, which causes those $B$ flares to be unrecorded. 
Therefore, we are training and testing the model with mislabeled data samples for predicting any class of flares, hence many weak flares that the model predicts probably are classified as false positives, which lowers the skill scores.

\subsection{Comparison with previous results}\label{subsec:comp_previous}
In this subsection, we will compare the results of this paper to previously published works. In the past decade, there have been several works that applied machine learning based models to predict solar flares. Those models including MLP (Multilayer perceptrons) \citep{2018SoPh..293...28F}, SVM (Support Vector Machine) \citep{2007SoPh..241..195Q, 2010RAA....10..785Y, 2015ApJ...798..135B, 2015ApJ...812...51B, 2015SpWea..13..778M, 2018SoPh..293...28F}, DeFN (Deep Flare Net) \citep{2018ApJ...858..113N} and a recently published work \citep{2019ApJ...877..121L} which also used LSTM network.

The skill scores for different models are shown in Table \ref{tab:comp_skill_scores}. Even though the input and the testing samples may be different in those works using MLP, SVM and DeFN models, the results are representative of the best performance those models can achieve on the solar flare prediction task. The results from the LSTM models outperform the MLP, SVM and DeFN models substantially, which indicates that taking time series data into account can substantially improve solar flare forecasting. As discussed in subsection \ref{subsec:diff_train_test_split}, we should use similar testing sets when ciging the various models since different train/test splitting can produce different skill scores. The notations for LSTM models are defined as follows: LSTM-15\_18 and LSTM-2015 are the models used in this paper. LSTM-15\_18 uses ARs from year 2015 to year 2018 for testing and LSTM-2015 uses ARs in year 2015 for testing. LSTM-Liu uses the model reported in \cite{2019ApJ...877..121L} which uses ARs in year 2015 for testing. 
Among those evaluation metrics, \textbf{1}. Accuracy (ACC) is the least useful since the data set is highly biased (flare events are rare so the majority samples in the testing set are negative), predicting those negative samples correctly leads to a high accuracy, however not much predictability for strong flare events may actually be achieved. 
\textbf{2}. Precision and POD reflect the model's ability of making positive predictions: Precision is the fraction of correctly predicted samples among all predicted positive samples. 
POD is the fraction of correctly predicted positives among the actual positive samples in the testing set. 
Therefore Precision provides more useful information about predictability of rare events, while POD by itself is not representative of the predictive skills. 
From Table \ref{tab:comp_skill_scores}, our model produces Precision scores within the same range as \cite{2019ApJ...877..121L}, and is better than the previous works listed.
\textbf{3}. HSS and TSS are often considered as good metrics for evaluating model predictability in binary classification tasks. 
However, TSS benefits from the large number of correctly predicted negative samples (TN) so that it is much higher than the HSS in all tests. 
HSS lessens the importance of true negatives (TN) but it is more sensitive to the fraction of positive samples in the testing set: the value of HSS will be higher if there are more positive samples in the testing set (which can be artificially achieved by creating a testing set that contains larger fraction of positive samples than the actual data). 
Our model gives better HSS than previous models (MLP, SVM and DeFN) that do not use time sequences and also has similar performance as the recently published results \citep{2019ApJ...877..121L} using LSTM, which validates the correctness of this work. \textbf{4}. For rare events,
small False Alarm Rates (FAR) are highly indicative of good prediction skills. 
The FAR for predicting $\ge C$ and $\ge M$ flares are all less than 0.1, which means that more than 90\% negative predictions from our model are correct. 
This contributes greatly to the high HSS values.

The details of our and the \cite{2019ApJ...877..121L} LSTM models are different but they obtain similar skill scores according to Table \ref{tab:comp_skill_scores}. This suggests that both LSTM models extract most of the useful information from the SHARP parameters and further improvement will require using more information from the observation. 

\begin{deluxetable}{cclll}
    \tablenum{3}
    \tablecaption{Comparison of skill scores for different models. Lines with * are results from this work. LSTM-2015 uses the same time period ase DeFN while LSTM-15\_18 uses the same time period as LSTM-Liu for the testing data set. \label{tab:comp_skill_scores}}
    \tablewidth{0pt}
    \tablehead{
    \colhead{Metric} & \colhead{Model} & \colhead{$\ge M$ class} & \colhead{$\ge C$ class} & \colhead{Any Class}
    }
    \startdata
    POD     & MLP       & 0.812  &  0.637  & -\\
            & SVM       & 0.692  &  0.746  & -\\
            & DeFN      & 0.891  &  0.761  & -\\
            & LSTM-Liu  & 0.881  &  0.762  & -\\
    *       & LSTM-2015 & 0.685  &  0.643  &  0.625\\
    *       & LSTM-15\_18 & 0.730 & 0.621  & 0.530 \\
    \hline
    Precision & MLP       & 0.143  &  0.451  & -\\
            & SVM       & 0.106  &  0.497  & -\\
            & DeFN      & 0.173  &  0.497  & -\\
            & LSTM-Liu  & 0.222  &  0.544  & -\\
    *       & LSTM-2015 & 0.311  &  0.677  &  0.670\\
    *       & LSTM-15\_18 & 0.282 & 0.635 & 0.702 \\
    \hline
    ACC     & MLP       & 0.855  &  0.778  & -\\
            & SVM       & 0.824  &  0.803  & -\\
            & DeFN      & 0.872  &  0.801  & -\\
            & LSTM-Liu  & 0.909  &  0.829  & -\\
    *       & LSTM-2015 & 0.929  &  0.858  &  0.814\\
    *       & LSTM-15\_18 & 0.945 & 0.883 & 0.800 \\
    \hline
    HSS     & MLP       & 0.204  &  0.389  & -\\
            & SVM       & 0.141  &  0.472  & -\\
            & DeFN      & 0.253  &  0.476  & -\\
            & LSTM-Liu  & 0.347  &  0.539  & -\\
    *       & LSTM-2015 & 0.394  &  0.567  & 0.519\\
    *       & LSTM-15\_18 & 0.382 & 0.557 & 0.473 \\
    \hline
    TSS     & MLP       & 0.669  &  0.449  & -\\
            & SVM       & 0.520  &  0.562  & -\\
            & DeFN      & 0.763  &  0.572  & -\\
            & LSTM-Liu  & 0.790  &  0.607  & -\\
    *       & LSTM-2015 & 0.623  &  0.559  &  0.509\\
    *       & LSTM-15\_18 & 0.681 & 0.553 & 0.439 \\
    \hline
    FAR     & MLP       & 0.143  &  0.188  & -\\
            & SVM       & 0.172  &  0.184  & -\\
            & DeFN      & 0.128  &  0.189  & -\\
            & LSTM-Liu  & 0.091  &  0.155  & -\\
    *       & LSTM-2015 & 0.062  &  0.084  &  0.116\\
    *       & LSTM-15\_18 & 0.049 & 0.068 & 0.092 \\
    \enddata
\end{deluxetable}

\subsection{Choosing different testing years}\label{subsec:diff_train_test_split}

\begin{figure}[!htb]
    \centering
    \includegraphics[width=0.45\textwidth]{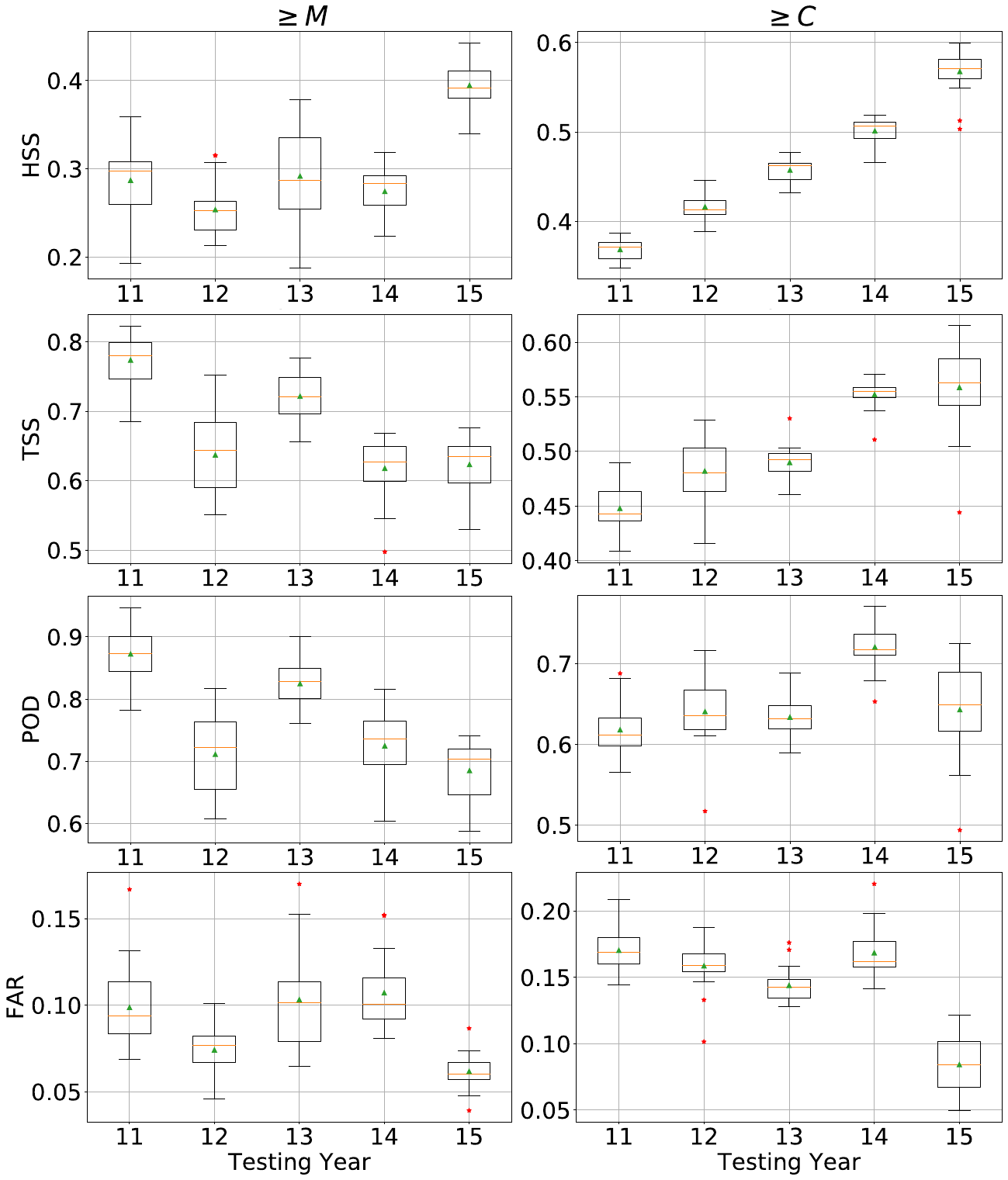}
    \caption{The box plots of four skill scores for different training and testing year choices in the 2011$-$2015 data set. Note that smaller FAR means better performance. The left and right columns show results for predicting flares of class $\ge$ M and $\ge$ C flares, respectively. The yellow line is the median and the green triangle is the mean of the 20 independent training runs. The lower and upper bounds of the boxes correspond to the quartiles, while the red stars are the data points outside the range of median $\pm 2.698 \sigma$. The mean value and median are calculated including the outliers.}\label{fig:comp_year}
\end{figure}

As described before in section \ref{subsec:train_test_split}, it is important to totally separate the training and testing samples. 
However, whether choosing different years of flares for testing can have different skill scores is still unclear since the previous works all used data before 2015 for training and after 2015 for testing. 
In this subsection, we conduct the training-testing process on different combinations of training and testing years. 
In Figure \ref{fig:comp_year}, we present the box plots of skill scores for twenty independent runs with the testing samples being one of the years from 2011 to 2015, and the other four years are used for training. 
(As shown by Table~\ref{tab:solar_flare_number}, there were very few large flares from 2016 to 2018, so those years are not suitable for testing and would not contribute much to training either.)

Figure \ref{fig:comp_year} shows that training on 2011 to 2014 and testing on 2015 gives the best HSS and FAR scores for predicting both $\ge C$ and $\ge M$ flares. The trend is different for the TSS for predicting $\ge M$ flares, because TSS is dominated by the POD values, but this does not mean truly good prediction for rare events, such as $\ge M$ flares. Good prediction of rare events requires very few false alarms and this will produce high HSS. Apparently, the model is quite "restrained" on making positive predictions for the year 2015 data, which improves its FAR and HSS scores. 
It is clear from Figure \ref{fig:comp_year} that evaluating the model performance on different years will introduce significant differences in the results.

\begin{figure}[htb!]
    \centering
    \includegraphics[width=0.45\textwidth]{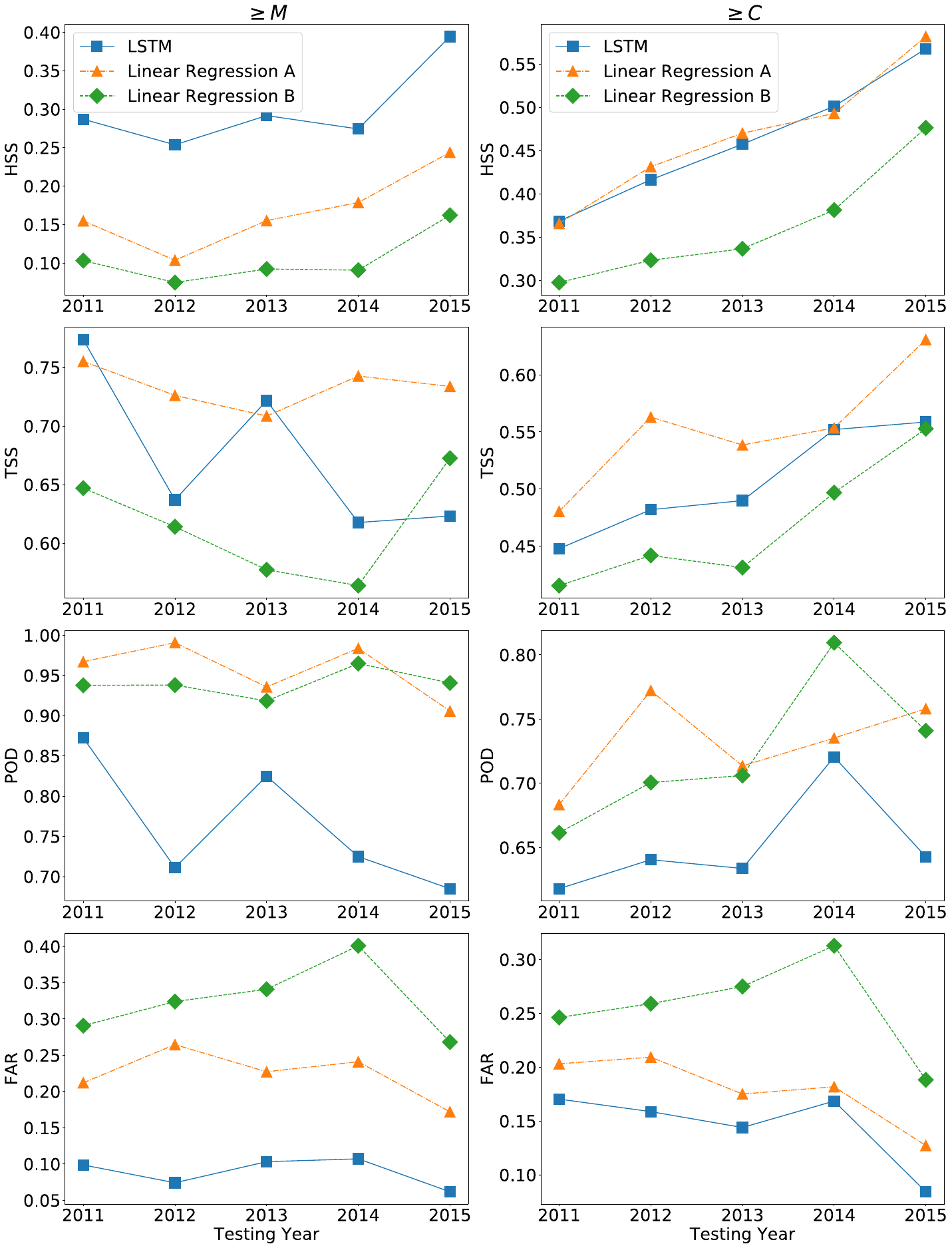}
    \caption{The comparison between LSTM model and two baseline linear regression models. Four skill scores are presented. Left column is the prediction for $\ge M$ flare and right column is for $\ge C$ flares. The Linear Regression A takes the whole time sequence as input (same as LSTM) and the Linear Regression B takes the mean value of the time sequence for each SHARP parameter. }\label{fig:comp_base_model}
\end{figure}

To investigate why the model produces fewer false alarms on year 2015 than other years, we set up two linear regression models as the baseline. 
These two baseline models use the same training and testing samples as the LSTM model. The time sequences of SHARP parameters are reshaped to one-dimensional vectors as the input of the first linear regression model, denoted by "Linear Regression A". 
For the second linear regression model, denoted by "Linear Regression B", the mean values of the time sequences are taken as input. Twenty independent runs are conducted and the mean values of the skill scores from the LSTM model and two baseline models are shown in Figure \ref{fig:comp_base_model}. The difference between the LSTM model and the Linear Regression A is the non-linearity introduced by the LSTM network. The Linear Regression B eliminates the time sequence information and only inputs the average level of activity into the model. From Figure \ref{fig:comp_base_model}: \textbf{1.} For predicting $\ge M$ flares, the LSTM gives the best HSS, followed by Linear Regression models A and B. For predicting $\ge C$ flares, the LSTM model has similar HSS as the Linear Regression A model and both are better than the Linear Regression B model. This illustrates the importance of the time sequence information. \textbf{2.} The linear regression models give larger POD and FAR than the LSTM. Therefore, the LSTM model has less tendency to make positive predictions, which results in better HSS. The optimal case is when the model produces high POD while also keeping a low FAR, which is not the case for an LSTM. This is the reason why an LSTM cannot provide a high TSS compared to Linear Regression models.

All three models give the lowest FAR when testing on year 2015 and for other testing years the thee models show similar trends. This indicates that the reason for the LSTM producing low FAR is related to the mean values of the SHARP parameters, which is the information used by the simplest Linear Regression B model. In conclusion, the LSTM model produces more reliable positive predictions than the simple linear regression models although it will miss some positive events. The model gives different results when being tested/trained on different years of data, apparently due to differences in the average SHARP parameters. 

\subsection{Training/Testing on different solar cycle phases}

According to subsection \ref{subsec:diff_train_test_split}, training and testing on different years leads to very different results. 
One possible reason for the difference could be the changes in the data processing procedure. However, from \cite{2018SoPh..293...45H}, although the data processing techniques were modified in January, 2015, those changes will not have major effects on the data products used in this work. 

To investigate if there are any intrinsic differences of the active regions from each year in the solar cycle, we first do the training and testing separately on each year. Since the number of active regions and flares is very small in years 2016 to 2018, those three years are grouped together. 
Four skill scores are collected to evaluate the model performance: HSS, TSS, POD and FAR. The process for selecting data samples is:
\begin{itemize}
    \item[1.] For active regions in each year, randomly select 25\% for testing and rest of them for training.
    \item[2.] Extract 48h time series of SHARP parameters for training and testing from active regions selected in step 1 and label the time series with the maximum flare class in the next 24 hours from the GOES flare record.
    \item[3.] Randomly drop negative samples in the training and testing sets to fix the ratio of positive samples and negative samples to be 0.05 for predicting $\ge M$ flares and 0.3 for predicting $\ge C$ flares.
\end{itemize}
Notice that in step 3, we fix the ratio of positive to negative samples to make the HSS directly comparable across different years. 
In addition, having a fixed positive/negative sample ratio makes the HSS and TSS behave similarly. 
We perform multiple runs with randomly dropping negative samples, so in fact the runs use different data sets. 
For each run, the skill scores are the mean values of the model outputs from the third to the tenth epochs. 
This range of epochs is chosen based on the typical evolution of skill with epochs: there is an initial rapid improvement, followed by a plateau with random oscillations, and finally worsening trends due to overfitting. 
The averaging over multiple epochs reduces the random variation due to the relatively small data sets.
In addition, the mean values better reflect the general performance of the model than picking the best epoch for each run.

\begin{figure}[htb!]
    \centering
    \includegraphics[width=0.45\textwidth]{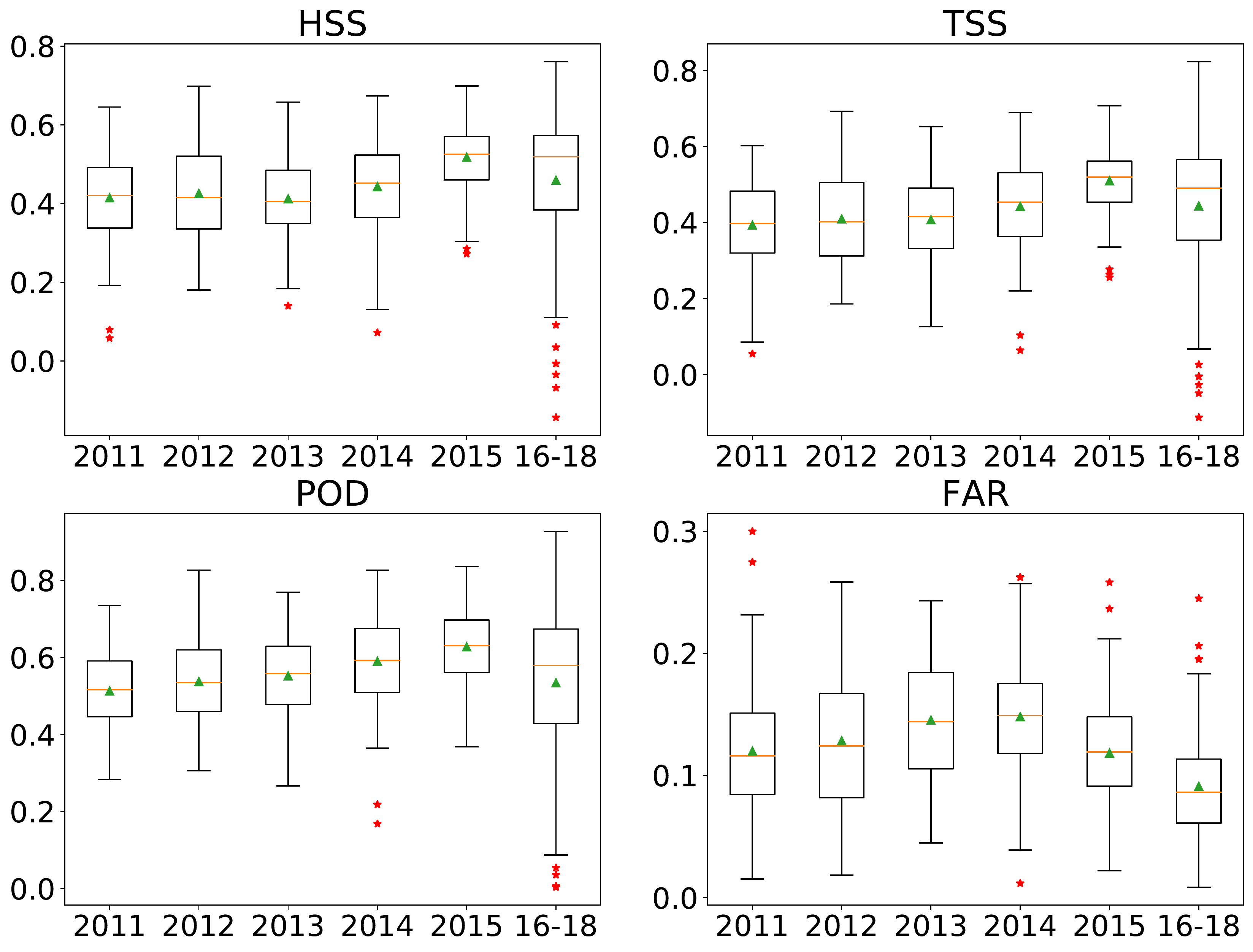}
    \caption{The box plots of skill scores for predicting $\ge C$ flares on different years from 2011 to 2018. The symbols in the figures are the same with previous box plots.}\label{fig:C_each_year}
\end{figure}

The box plots of skill scores for predicting $\ge C$ flares are shown in Figure \ref{fig:C_each_year}. 
Each box contains 100 data points from 100 runs using randomly selected active regions for training/testing and dropping negative samples in the data sets. 
Because there are few active regions and flare events during the three years from 2016 to 2018 (see Table \ref{tab:solar_flare_number}), the skills scores are less centered and the number of outliers is larger than those from other years. 
The results show that training and testing on the data after 2015 produces better skill scores than the earlier years. The FAR has the most substantial difference for data sets after and before 2015, which is also the major reason for better HSS and TSS since the POD does not vary much. 
We are not showing results for predicting $\ge M$ flares on each year separately because the $\ge M$ flares are too rare to give any statistically significant results on such small data samples.

\begin{figure}[htb!]
    \centering
    \includegraphics[width=0.45\textwidth]{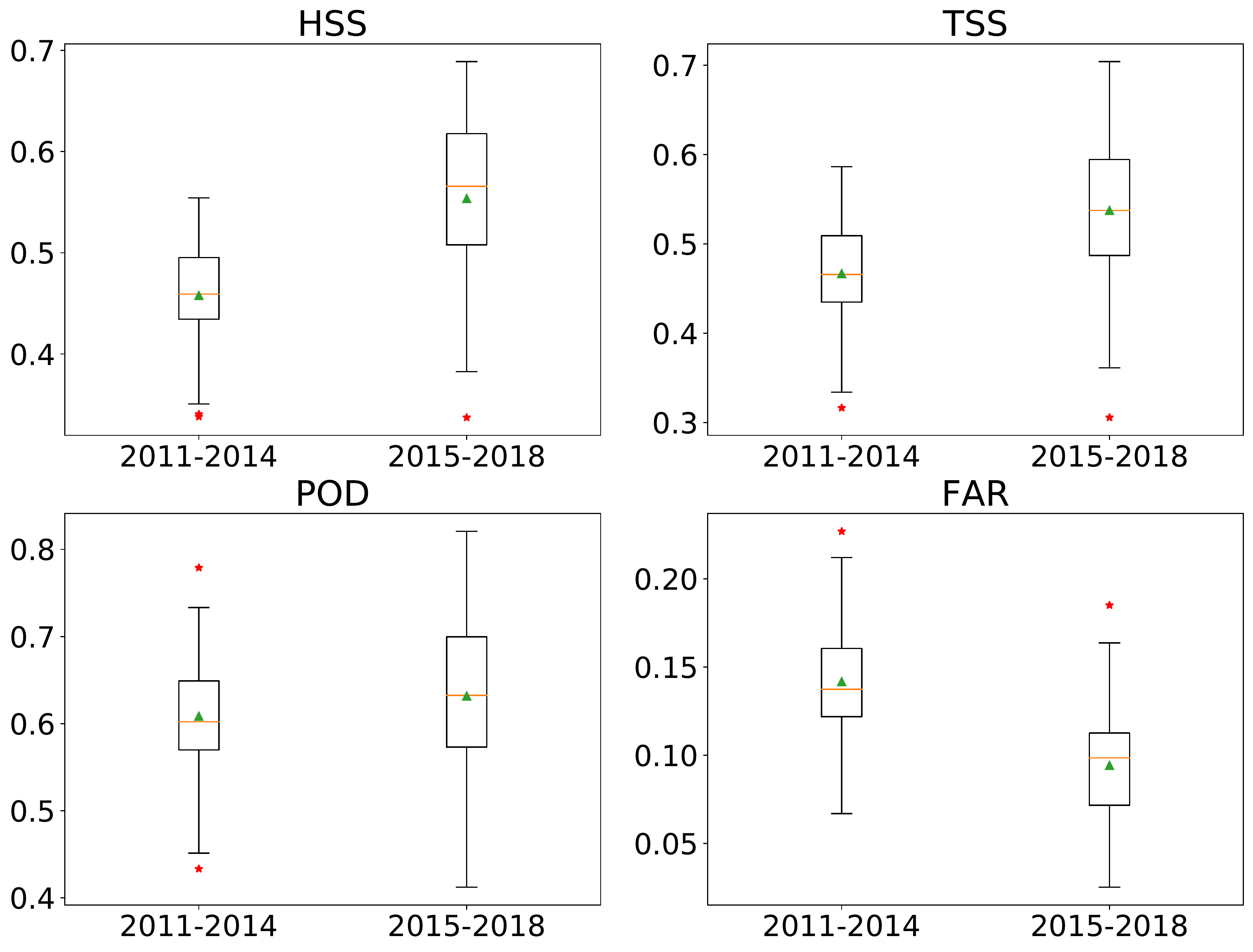}
    \caption{The box plots of skill scores for predicting $\ge C$ flares. The training/testing are conducted within one of two time periods: 2011-2014 and 2015-2018. The symbols in the figure are the same as in previous box plots.}\label{fig:C_two_periods}
\end{figure}

\begin{figure}[htb!]
    \centering
    \includegraphics[width=0.45\textwidth]{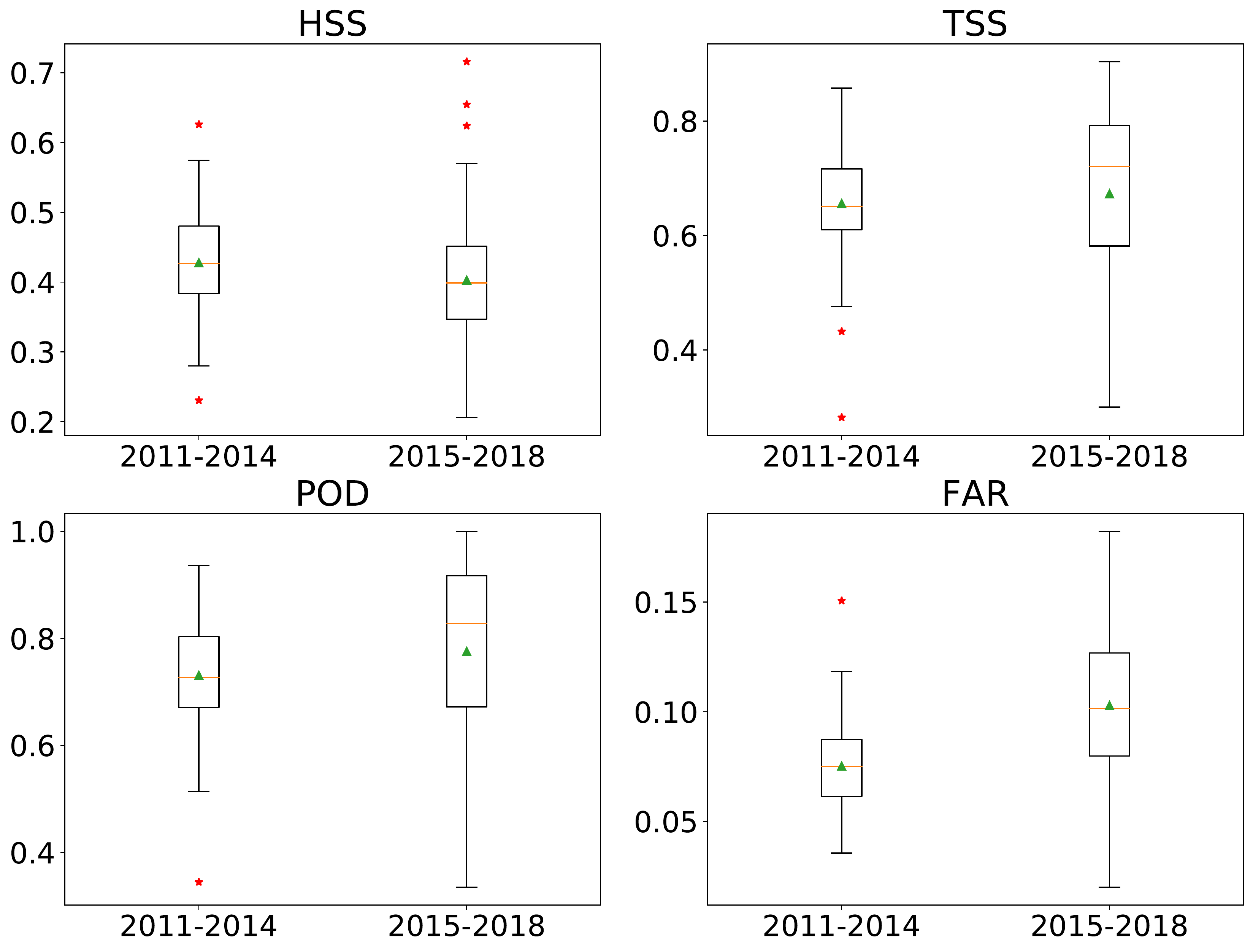}
    \caption{The box plots of skill scores for predicting $\ge M$ flares. The training/testing are conducted within one of two time periods: 2011-2014 and 2015-2018.. The symbols in the figure are the same as in previous box plots.}\label{fig:MX_two_periods}
\end{figure}

The results in Figure \ref{fig:C_each_year} show that the prediction model has better performances for the years 2015-2018 than for 2011-2014. 
To demonstrate this difference with better statistics, we conduct the training/testing process on two data sets using these two time periods, and compare the results.
The data selecting strategy is the same as what we did for training/testing on each year but now there are enough M and X flares to produce statistically robust results. 
The skill scores for predicting $\ge C$ and $\ge M$ flares are shown in Figures~\ref{fig:C_two_periods} and \ref{fig:MX_two_periods}, respectively. For predicting $\ge C$ flares, the HSS, TSS scores are clearly and significantly better for 2015-2018 than for 2011-2014. 
The box plots of POD overlap with each other while the box plots of FAR are well separated. 
This result indicates that for capturing $\ge C$ flares in the testing set, the model has similar performance when trained on the data samples from two phases in a solar cycle. 
However, the model will produce much fewer false alarms if it is trained on data samples from the declining phase in a solar cycle. 
For predicting $\ge M$ flares, according to Figure \ref{fig:MX_two_periods}, the difference of skill scores is less obvious. 
In general, the results from 2015-2018 show larger variance because there are fewer $\ge M$ flares in this time range. The box plots of HSS, TSS and POD overlap to a large extent except the model gives a higher FAR when trained/tested on year 2015-2018. 

\begin{figure}[htb!]
    \centering
    \includegraphics[width=0.45\textwidth]{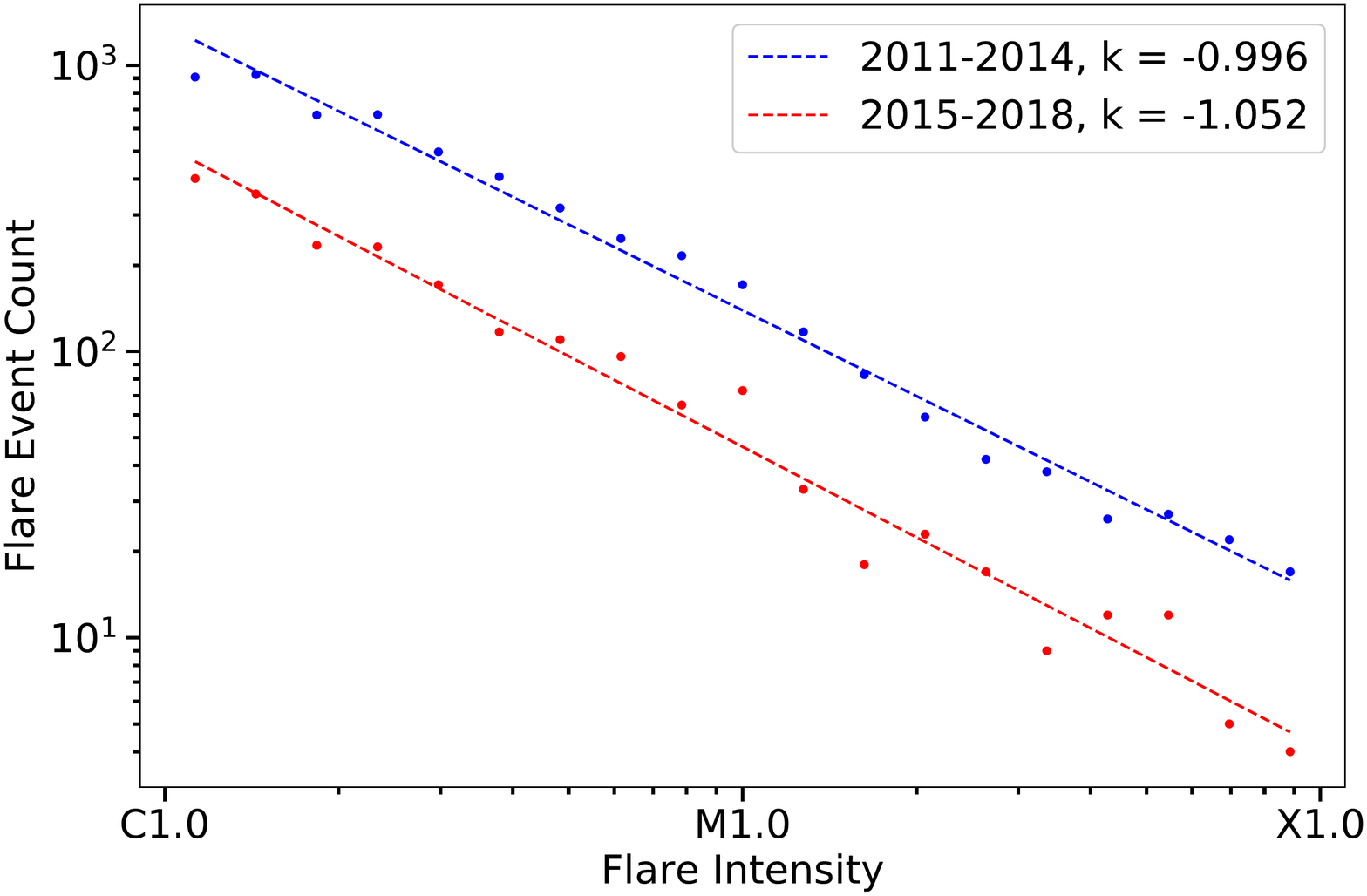}
    \caption{The histogram for C and M flares in two groups of years. The plot is in log-log scale and the histograms are straight lines which shows the intensity of flares agrees with the power-law distribution.}\label{fig:flare_count_hist}
\end{figure}

There is a significant difference between 2011-2014 and 2015-2018 for predicting $\ge C$ flares, but not for $\ge M$ flares. 
A simple explanation could be different flare intensity distributions in two phases of a solar cycle. 
For example, if the frequency of $C$ and $M$ flares are better separated from each other in one of the time periods (i.e. there are relatively fewer flares with energies near the C/M class boundary) then it will help the model to correctly identify $\ge$ M flares. 
To check this possibility we show the distribution of flare events as a function of energy on a log-log scale for the two time periods in Figure \ref{fig:flare_count_hist}. 
The histograms are well approximated with straight lines, indicating that the flare intensity distributions approximately follow power laws \citep{lu1991avalanches}.
While the amplitudes are different by about a factor of 3 (there are more flares during solar max than during solar min), the slopes of the two lines are very close to each other. 
So there is no obvious difference in the shape of the flare intensity distributions between the two phases of a solar cycle. 
There seems to be some excess of flares near the C/M class boundary for 2015-2018, which may contribute to the worse performance on predicting $\ge$ M flares than $\ge$ C flares for this time period.
However, we set the threshold of positive and negative classes to C8.0. All these excess of flares are labeled as positive. The results show no difference in skill scores for two time periods.
Thus we conclude that difference of the skill scores in Figure \ref{fig:C_two_periods} and \ref{fig:MX_two_periods} cannot be simply explained by different flare intensity distributions in the two time periods.

\section{Conclusion}\label{sec:discussion_conclusion}

In this paper, we build a data set covering active regions from 2011 to 2018 from \href{http://jsoc.stanford.edu}{Joint Science Operations Center} (\texttt{jsoc.stanford.edu}). Each data sample is the time sequence of twenty SHARP parameters, which represent the magnetic field properties of an active region. We develop an LSTM network to predict the maximum flare class $\Gamma$ in the next 24 hours produced by an active region. The prediction task is reduced to a binary classification when $\Gamma$ is a combination of classes above a certain threshold. We consider three different cases for $\Gamma$: $\ge M$, $\ge C$ and $\ge A$. The last case corresponds to predicting any flares. The training/testing splitting is based on active regions, which guarantees that the model is tested on data samples that it has never seen previously. The skill scores produced by the model vary substantially for different years and we investigate the solar cycle dependence of the model performance. The main results of this paper are summarized as follows:

\begin{enumerate}
    \item For evaluating models predicting rare events, such as solar flares, the most relevant skill score is the Heidke Skill Score (HSS) that is strongly correlated with low false alarm rate (FAR). On the other hand, HSS is sensitive to the ratio of positive and negative samples in the testing set, which means that comparison of model performance for different data sets requires caution.
    
    \item  The LSTM based model achieves better HSS for predicting solar flares than the previous approaches such as MLP, SVM and DeFN. Using the time series information improves relevant skills. Our results are also comparable with the recently published work using a similar LSTM method.
    
    \item Although more than 50\% percent of skill scores of LSTM model can be acquired from simple linear regression models, the non-linearity introduced by LSTM reduces the number of false alarms and improves the prediction skills of the model.  
    
    \item Previous works using active region data after 2015 for testing could introduce bias into the skill scores. If the model is trained on 2011-2014 and tested on 2015, it produces better skill scores than other combinations of training and testing years. This appears to be related to the difference of average level of solar activity in the training and testing sets.

\end{enumerate}

Based on the results presented in this paper, the LSTM is a valid method for the solar flare prediction task. The skill scores from this paper are very close to those generated by other different LSTM models indicates that the information contained in the SHARP parameters is limited. In future work, we plan to use more observational information to further improve the flare prediction skills.

\newpage

\bibliography{main}{}
\bibliographystyle{aasjournal}



\end{document}